\documentclass[%
 reprint,
 amsmath,
 amssymb,
 aps,
]{revtex4-2}

\usepackage{graphicx}
\usepackage{dcolumn}
\usepackage{bm}
\usepackage{comment}
\usepackage[mathlines]{lineno}
\usepackage{physics}
\usepackage[colorlinks = true,
            linkcolor = blue,
            urlcolor  = blue,
            citecolor = blue,
            anchorcolor = blue]{hyperref}

\usepackage{color}
\usepackage[dvipsnames]{xcolor}
\usepackage[normalem]{ulem}

\begin{document}
\preprint{APS/123-QED}

\title{Single Photon Cross-Phase Shifts Can Be Enhanced by Localization in both Frequency and Time}
\author{Andy Jiao}
\thanks{These authors contributed equally to this work.}
\author{Vida-Michelle Nixon}
\thanks{These authors contributed equally to this work.}
\author{Kyle Thompson}
\author{Aephraim Steinberg}
\affiliation{Department of Physics, University of Toronto, Toronto, Ontario M5S 1A7, Canada}

\begin{abstract}
Single-photon optical nonlinearities face a fundamental trade-off: maximum nonlinearity requires both spectral resonance (narrow bandwidth) and high peak intensity (short duration), constraints that are incompatible due to the time-energy uncertainty relation. 
We demonstrate experimentally that this limitation does not need to exist in cases involving post-selection.
We measure a cross-phase shift (XPS) produced by a resonant photon from a narrow-band source that is first transmitted through a cold atomic cloud and then localized in time through detection. 
The peak size of this XPS is greatly enhanced compared to that of Gaussian single-photon-level pulses without post-selection, benefiting from the narrow bandwidth of the resonant prepared state and the high intensity of the post-selected state simultaneously. 
We measure enhancements in the peak XPS of 6$\pm$1 at an optical depth (OD) of 2.4$\pm$0.1, and our results are in qualitative agreement across a range of optical depths with the recently developed weak value theory of atomic excitation [Thompson et al., APL Quantum 2, 036108 (2025)] for such post-selected photons. 
This work uncovers new consequences of having simultaneous knowledge of frequency and time, raising new foundational questions about how a particle behaves, and interacts with other systems, when its preparation and post-selection are non-commuting.
\end{abstract}

\maketitle

\textit{Introduction - }What happens when we let a narrow-band photon interact with another system and then observe it at a particular moment in time, better defined than its initial coherence time? Should we think of the interaction as having been tightly localized in time?
Optical nonlinearities, of which cross-phase shifts (XPS) are an example, are strongest on resonance, but grow with intensity (and hence temporal localization). The peak XPS one photon can impress on a “probe” via an atomic nonlinearity has a maximum for a temporal duration on the order of the atomic lifetime, falling at higher bandwidths because of the loss of resonance, and at longer durations because of the reduced peak intensity.
However, studies of post-selected systems \cite{aharonov1988} have shown that if a system is prepared in an eigenstate of one observable and later detected in an eigenstate of another (even incompatible) one, the effects of this system on any others to which it is weakly coupled can benefit from both eigenvalues simultaneously. This motivated us to perform an experiment in which a photon is prepared in a very narrow-band state tuned to resonance, but detected with sub-lifetime resolution, to test whether its nonlinear effect on a second optical beam can simultaneously benefit from the frequency-localization of the preparation and the time-localization of the post-selection.

While Heisenberg's uncertainty relation and measurement-disturbance relations are concerned with knowing two incompatible observables of a single quantum system, this work in contrast is about the effect (on a different system) of a particle whose prepared state and post-selected state have precise values of frequency and time respectively. When the frequency and time of a photon are simultaneously known in this way, the photon can share qualities of eigenvalues in both bases, leading to potentially large peak effects by the photon. There has been a great amount of work on producing large single photon nonlinearities \cite{Turchette1995, Fushman2008, Matsuda2009, Lo2011, Shiau2011, Venkataraman2013, Beck2014, Feizpour2015, Tiarks2016, Beck2016, Hosseini2016, Hallaji2017, Duan2020} for the purpose of making two-qubit gates for photons \cite{Milburn1989, Nemoto2004}. Some of these implementations also use weak values to realize or enhance the nonlinearity \cite{Feizpour2015, Hosseini2016, Hallaji2017, Duan2020}. Weak value amplification, a technique in which the weak value is enhanced by using a pair of prepared and post-selected states whose inner product is small, is one such approach \cite{Feizpour2011, Hallaji2017}. The peak nonlinearity in this work can in fact also be analyzed through the lens of weak value amplification. Previous works have shown that the cross-phase nonlinearity here is proportional to the weak value of the atomic excitation caused by the photon \cite{Sinclair2022, Thompson2025, angulo2026}. One can rigorously show that the peak of this weak value is 1-exp(OD/2) when the photon is prepared resonantly and is post-selected to be a delta function in time in the transmission. (A derivation is in the Appendix.) While there has been considerable interest in how much atomic excitation a single photon can cause by shaping the photon temporally \cite{Sondermann2007, Wang2011, Golla2012, Aljunid2013, Leong2016, Rag2017}, a XPS produced by a single photon here even at a moderate OD would correspond to several atomic excitations in magnitude. 

\enlargethispage{\baselineskip}

\begin{figure*}[t]
   \includegraphics[width = 0.7\linewidth]{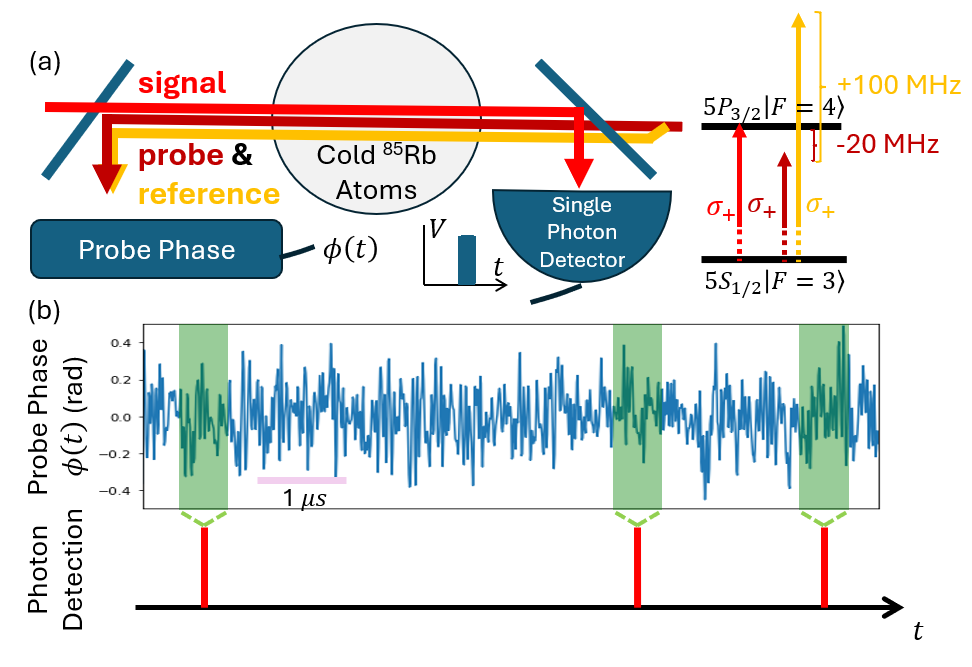}
   \caption{ 
   a) \textbf{Experimental Schematic} Photons in the continuous-wave resonant signal produce cross-phase shifts (XPS) on the probe in the cloud of atoms. The probe phase is measured by interfering the probe with the reference and measuring the phase of the beat note.
   b) \textbf{Data Analysis Procedure} Probe phase measurements (top) are windowed around the signal photon detections (bottom) and averaged. The probe phase measurements shown are actual sample data. (The duration of the windows is not to scale.)} 
   \label{schematic}
\end{figure*}

\textit{Experiment Description} - The experimental setup is shown schematically in Fig. \ref{schematic}a. We use the 5S\textsubscript{1/2}$\ket{F=3}$ and 5P\textsubscript{3/2}$\ket{F=4}$ states of \textsuperscript{85}Rb as the two-level system. Both the 780 nm signal and probe beams address this transition. The signal beam photons excite the atoms and the probe beam senses the amount of excitation as a phase shift. The transition has a lifetime of 26 ns and a linewidth of 6.1 MHz. The signal and probe beams have linewidths less than 1 MHz. 
The experimental duty cycle consists of 2 ms of optical trapping and optical molasses cooling, followed by a 1 ms period of data collection. The atoms are cooled to around 40 $\mu$K, and the atomic cloud is $\sim$ 1 mm in diameter. 
The experiment is done at a range of optical depths from 1.0 to 4.7.
We measure the optical depth (OD) by scanning the probe through resonance and measuring the transmission profile at the beginning of the data collection period.
During the data collection period, the resonant CW signal beam is turned on and has a power of $\sim$ 0.05 nW. It has a beam waist radius of approximately 25 $\mu$m at the atoms, resulting in a peak intensity of around 0.005 mW/cm$^2$, far below the saturation intensity of the transition. The transmitted photons in the signal beam are detected using a single photon detector. The rate of detection is kept around 0.6 photons per $\mu s$. 
The probe beam and the signal beam are spatially overlapped but counter-propagating (so that they can be separated after interaction).
The probe is detuned from resonance by about -20 MHz and  experiences a change in its phase proportional to the atomic excitation. The probe has a co-propagating sideband that is +100 MHz frequency-shifted, serving as its phase reference. The probe has a power of around 2 nW while the sideband has a power around 8 nW. 
To extract the phase shift of the probe, we measure the phase shift of the beat note between the probe and its sideband via radio-frequency heterodyne, as done in our previous experiments \cite{angulo2026, Sinclair2022, Hallaji2017, Feizpour2015}. We detect the combined probe and its +100 MHz sideband using an AC-coupled high-bandwidth photo detector to obtain the beat note. The beat note is then sent into an IQ demodulator. (The local oscillator of the IQ demodulation is roughly 4 kHz detuned from the beat note frequency.) We used 25 MHz low-pass filters in the I and Q ports of the IQ demodulator before digitizing the voltage readings at 8 ns resolution. 

When we detect a photon from the transmitted signal beam, we window the probe phase measurements around the time of detection of the photon, as shown in Fig \ref{schematic}b. The quantity being measured here is the atomic excitation produced in the cloud of atoms by the single post-selected photon, with the atomic excitation proportional to the XPS on the probe. In the weak measurement framework, our measurement gives the weak value of the atomic excitation produced by a single photon prepared in the resonant continuous-wave state and then post-selected in a temporal mode after transmission. This is because the single photon weak value can be obtained from a coherent state by conditioning the measurement of an observable on the detection of a single photon. For a formal proof, see Ref. \cite{Wiseman2023}. Additional information on this technique as applied to this experiment is given in the Appendix. 

\begin{figure*}[t]
  \centering
  \includegraphics[width=0.9\textwidth]{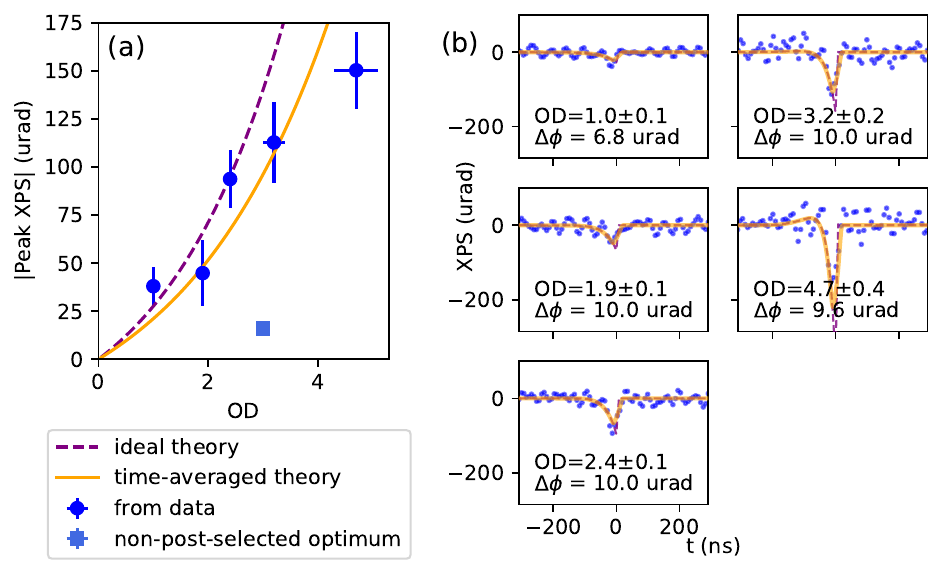}
  \caption{(a) Magnitude of the peak value of the measured XPS on the probe from the single photon. The OD values are those labeled in Fig. (b), which are OD values measured via the probe frequency scan. The theory is plotted assuming the signal OD is the same as the probe OD. In (a), the non-post-selected optimum is the optimum in Fig. \ref{peak_vs_pulsewidth}. (b) XPSs due to the single photon observed across different ODs. In (b), $\Delta \phi$ is the phase measurement uncertainty after averaging. The theory predictions for atomic excitation are scaled by a factor of 47 urad / atomic excitation to fit the experimental observations. The ideal theory (purple dashed) convolved with a rectangle with a 28 ns duration gives the time-averaged theory (in orange solid). To center in window, all data in (b) are shifted temporally by the same amount. The amount was determined by individual fits of the theory to the data, showing temporal alignment to within a few nanoseconds.}
  \label{fig:vs_OD_Data}
\end{figure*}

The details of the data analysis to obtain the plots in Fig. \ref{fig:vs_OD_Data} are as follows. Around the time that a photon is registered, we take a window of 4.8 us of probe phase measurements. We perform a linear fit and do a subtraction of it to remove the phase drift resulting from the 4 kHz detuning between the beat note and the local oscillator in our IQ demodulation. Many such windows of the linear background subtracted probe phase measurements are averaged together to reveal the feature of interest --- the XPS due to the single photon detected. Because the technique of this work uses correlations between the probe phase and signal detection probability, other sources of correlations between the two could generate additional features. After averaging, there is a slow oscillation around 300 kHz. Due to the low frequency, this oscillation does not affect the feature of interest. (The feature of interest always occurs near the peak of this slow oscillation, where this slow oscillation is approximately flat.) We remove this slow background by masking the feature and performing a sinusoidal fit, and then subtracting this fitted background. The result, windowed around the feature of interest, is what is shown in Fig 2b). For more information on this slow oscillation in the background, see the Supplemental Materials. Around the feature of interest, there is also a fast oscillation whose frequency is close to the probe's detuning. We document the behavior of this fast oscillation more in the Supplemental Materials. Since this fast oscillation occurs at the probe's detuning, it could be related to  \cite{Carmichael2000,Foster2000,Foster2002,Masters2023,Wang2025}, and is currently under investigation. To see this fast oscillation more clearly, refer to Supplemental Materials.

\textit{Observation of the Cross-Phase Shifts} - We observed the (single-photon) XPS across a range of ODs. Fig. \ref{fig:vs_OD_Data}a) summarizes the peak's dependence on OD. The OD is measured using the probe, and can differ slightly from the OD for the signal beam (which is the only relevant parameter in the theoretical predictions); there are also some potential systematic errors in the OD measurement, described further in the Supplemental Materials. The XPSs are individually shown in Fig. \ref{fig:vs_OD_Data}b). (The uncertainties of peak XPS in Fig. \ref{fig:vs_OD_Data} include both the phase measurement uncertainty and the spread of the fast oscillations on the wings of the data traces. See the Supplemental Materials for error analysis.) We see a peak XPS as big as -148$\pm$20 urad, large compared to our previous experiments, at an OD of 4.7$\pm$0.4. We make theoretical predictions for the XPSs using the theoretical framework in Ref. \cite{Thompson2025}. Our simulation specifies a constant wavefunction (before the atomic cloud) as the forward-evolving state and an 8 ns long rectangular wavefunction (after the atomic cloud) as the backward-evolving state. Since our theoretical framework makes a prediction only for atomic excitation, we compare with experimental observations of XPS by applying a scale factor of 47 $\mu$rad / atomic excitation, determined phenomenologically to match our measurements well.
The growth of XPS with OD in Fig \ref{fig:vs_OD_Data}a) is roughly consistent with our predictions, despite an undershoot at the highest OD we measure (possibly explained in part by the systematics in signal OD determination and day-to-day variations in the coupling of the probe beam to the excitations), but cannot firmly establish the exponential dependence without further measurements, at higher ODs.
When looking at the individual data traces in Fig. \ref{fig:vs_OD_Data}b), the agreement between theory and the data traces is satisfactory. However, the experimental traces lack the ``sharpness'' of the theoretical predictions (shown in purple). The data appear to be more consistent with the ideal theory smoothed over some time interval (shown in orange). A possible cause for the smoothing could be the temporal response of the IQ demodulation and the 25 MHz low pass filter used for the probe phase measurement. We separately characterized the response of the IQ demodulation set-up and saw a rise time of around 28 ns. Thus, we average the theoretical simulation over a 28 ns time interval for comparison. These are shown by the orange curves in Fig. \ref{fig:vs_OD_Data}b).

\begin{figure}[h]
   \includegraphics[width = \linewidth]{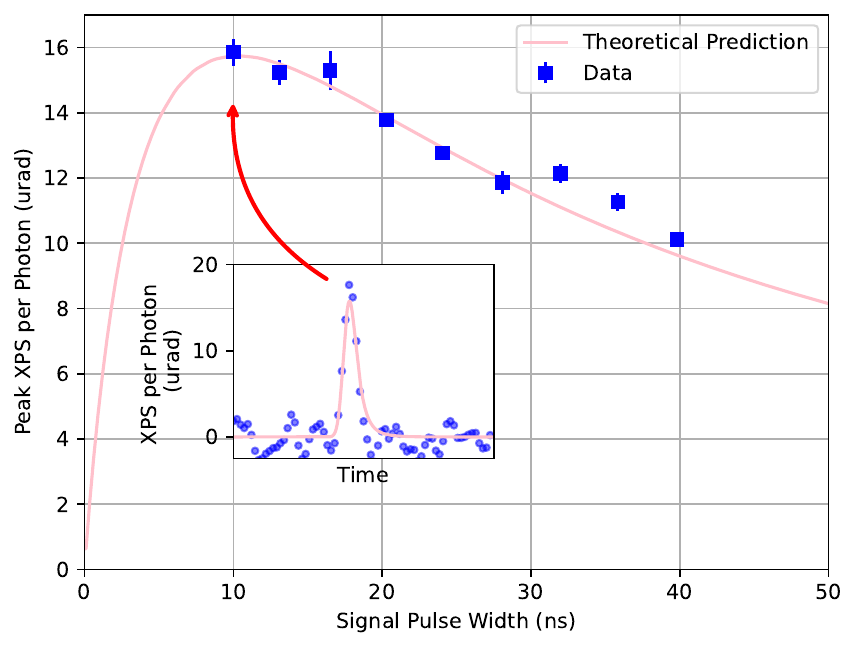}
   \caption{(Blue square) Peak XPS per photon vs pulse intensity rms for a pulsed input taken at OD = 2.9$\pm$0.1. (Pink solid line) Theoretical prediction for atomic excitation are converted using 47 urad / atomic excitation, which is the same as used for the post-selected data. Inset: example of XPS per photon at pulse width = 10 ns. }
   \label{peak_vs_pulsewidth}
\end{figure}

\textit{Comparison to Non-Post-Selected Case} - The fact that such a post-selected single photon can produce significantly higher atomic excitation than previously observed in similar experiments invites the question: How does one optimally excite an atom? Using the same theoretical framework as Ref. \cite{Thompson2025} we can make predictions for the atomic excitation produced by an incident single photon without post-selection. For such incident single photons without post-selection, there exists an optimal pulse duration for producing the maximal peak atomic excitation due to the trade-off between the bandwidth and duration described previously. Fig.\ref{peak_vs_pulsewidth} shows the peak XPS as a function of the duration of Gaussian pulses both experimentally measured and theoretically predicted. While the optimal pulse shape to use for generating atomic excitation is a rising exponential \cite{Sondermann2007,Wang2011,Golla2012,Aljunid2013,Leong2016,Rag2017}, we used Gaussian pulses due to experimental limitations. (According to Ref. \cite{Wang2011}, the peak produced by an optimized Gaussian pulse will only fall short of the rising-exponential optimum by a factor of 0.8.) The Gaussian pulses are generated with an acousto-optic modulator, through which the signal beam passes before heading to the atomic cloud. We generate one pulse every 576 ns (a ``shot"), during the 1 ms long data collection portion of our experimental duty cycle. Regardless of whether the detector fires or not, we examine the phase profile of the probe beam, and we divide the average of all 576 ns shots taken by the mean photon number in a pulse. When comparing the optimal value taken at OD=2.9$\pm$0.1 to the peak phase shift of the post-selected data (taken at an OD of 2.4$\pm$0.1), we observe an experimental enhancement in the peak value by at least a factor of $6\pm1$ at this OD. 

\textit{Conclusion} - We observed the XPS from a single photon which is prepared on resonance and then localized in time after transmission. There is both qualitative and rough quantitative agreement between the experimental data and the theoretical prediction across different ODs. We showed that the preparation and post-selection enhances the peak effect compared to what is achievable with single photon pulses without post-selection, observing a factor of 6$\pm$1 enhancement at an OD of 2.4$\pm$0.1. However, one should caution against an overly literal interpretation of the enhancement since it involves post-selection. For example, another consequence of the post-selection is that the atomic excitation corresponding to the peak phase shift of the post-selected data is in fact negative (i.e. the phase shift has an opposite sign to the non-post-selected data)! Nevertheless, the predicted exponential scaling of the peak size with OD makes this XPS a promising avenue for further investigation. In conclusion, by taking advantage of both the frequency and duration of a photon --- preparing it in the frequency of the strongest interaction and post-selecting it to be in a short duration --- it is shown that its peak effect on other systems can be beyond what the classical trade-off allows. This opens a new chapter in the study of joint measurement of non-commuting observables, and suggests a novel sense in which (post-selected) particles may benefit simultaneously from the determination of two incompatible properties.

Acknowledgements - This work was supported by NSERC Discovery Grant No. RGPIN-2020-05767, the John Templeton Foundation grant ID 63209, the QuEnSi quantum alliance (NSERC ALLRP 578468-22), and the Fetzer Franklin Fund of the John E. Fetzer Memorial Trust. We thank Mikhail Mamaev and Cecilia Soroco for helpful discussion. We thank Daniela Angulo for prior contributions to the experimental setup.

\bibliography{bibliography}

@article{Thompson2025,
  title = {How much time does a photon spend as an atomic excitation before being transmitted through a cloud of atoms?},
  author = {Kyle Thompson and Kehui Li and Daniela Angulo and Vida-Michelle Nixon and Josiah Sinclair and Amal Vijayalekshmi Sivakumar and Howard M. Wiseman and Aephraim M. Steinberg},
  journal = {APL Quantum},
  volume = {2},
  number = {3},
  pages = {036108},
  year = {2025},
  publisher = {AIP Publishing},
  doi = {10.1063/5.0288743}
}

@article{aharonov1988,
	title = {How the result of a measurement of a component of the spin of a spin- \textit{1/2} particle can turn out to be 100},
	volume = {60},
	copyright = {http://link.aps.org/licenses/aps-default-license},
	issn = {0031-9007},
	url = {https://link.aps.org/doi/10.1103/PhysRevLett.60.1351},
	doi = {10.1103/PhysRevLett.60.1351},
	number = {14},
	urldate = {2024-07-04},
	journal = {Physical Review Letters},
	author = {Aharonov, Yakir and Albert, David Z. and Vaidman, Lev},
	month = apr,
	year = {1988},
	pages = {1351--1354},
}

@article{Turchette1995,
  author    = {Q. A. Turchette and C. J. Hood and W. Lange and H. Mabuchi and H. J. Kimble},
  title     = {Measurement of conditional phase shifts for quantum logic},
  journal   = {Physical Review Letters},
  volume    = {75},
  pages     = {4710--4713},
  year      = {1995},
  doi       = {10.1103/PhysRevLett.75.4710},
}

@article{Fushman2008,
  author    = {I. Fushman and D. Englund and A. Faraon and N. Stoltz and P. Petroff and J. Vu{\v{c}}kovi{\'c}},
  title     = {Controlled phase shifts with a single quantum dot},
  journal   = {Science},
  volume    = {320},
  pages     = {769--772},
  year      = {2008},
  doi       = {10.1126/science.1154643},
}

@article{Matsuda2009,
  title = {Observation of optical-fibre Kerr nonlinearity at the single-photon level},
  author = {Matsuda, Nobuyuki and Shimizu, Ryo and Mitsumori, Yoshihiro and Kosaka, Hideo and Edamatsu, Keiichi},
  journal = {Nature Photonics},
  volume = {3},
  number = {2},
  pages = {95--98},
  year = {2009},
  publisher = {Nature Publishing Group},
  doi = {10.1038/nphoton.2008.292}
}

@article{Lo2011,
  author    = {H.-Y. Lo and Y.-C. Chen and P.-C. Su and H.-C. Chen and J.-X. Chen and Y.-C. Chen and I. A. Yu and Y.-F. Chen},
  title     = {Electromagnetically-induced-transparency-based cross-phase-modulation at attojoule levels},
  journal   = {Physical Review A},
  volume    = {83},
  pages     = {041804},
  year      = {2011},
  doi       = {10.1103/PhysRevA.83.041804},
}

@article{Shiau2011,
  author    = {B.-W. Shiau and M.-C. Wu and C.-C. Lin and Y.-C. Chen},
  title     = {Low-light-level cross-phase modulation with double slow light pulses},
  journal   = {Physical Review Letters},
  volume    = {106},
  pages     = {193006},
  year      = {2011},
  doi       = {10.1103/PhysRevLett.106.193006},
}

@article{Venkataraman2013,
  author    = {V. Venkataraman and K. Saha and A. L. Gaeta},
  title     = {Phase modulation at the few-photon level for weak-nonlinearity-based quantum computing},
  journal   = {Nature Photonics},
  volume    = {7},
  pages     = {138--141},
  year      = {2013},
  doi       = {10.1038/nphoton.2012.283},
}

@article{Beck2014,
  title = {Cross Modulation of Two Laser Beams at the Individual-Photon Level},
  author = {Beck, Kristin M. and Chen, Wenlan and Lin, Qian and Gullans, Michael J. and Lukin, Mikhail D. and Vuleti{\'c}, Vladan},
  journal = {Physical Review Letters},
  volume = {113},
  number = {11},
  pages = {113603},
  year = {2014},
  doi = {10.1103/PhysRevLett.113.113603},
  publisher = {American Physical Society}
}

@article{Feizpour2015,
  title = {Observation of the nonlinear phase shift due to single post-selected photons},
  author = {Feizpour, Amir and Hallaji, Mostafa and Dmochowski, Gregor and Steinberg, Aephraim M.},
  journal = {Nature Physics},
  volume = {11},
  number = {11},
  pages = {905--909},
  year = {2015},
  publisher = {Nature Publishing Group},
  doi = {10.1038/nphys3433}
}

@article{Tiarks2016,
  title = {Optical $\pi$ phase shift created with a single-photon pulse},
  author = {Tiarks, Daniel and Reiserer, Andreas and Ritter, Stefan and Rempe, Gerhard},
  journal = {Science Advances},
  volume = {2},
  number = {4},
  pages = {e1600036},
  year = {2016},
  publisher = {American Association for the Advancement of Science},
  doi = {10.1126/sciadv.1600036}
}

@article{Beck2016,
  title        = {Large conditional single-photon cross-phase modulation},
  author       = {Beck, Kristin M. and Hosseini, Mahdi and Duan, Yiheng and Vuletić, Vladan},
  journal      = {Proceedings of the National Academy of Sciences of the United States of America},
  volume       = {113},
  number       = {35},
  pages        = {9740--9744},
  year         = {2016},
  publisher    = {National Academy of Sciences},
  doi          = {10.1073/pnas.1524117113}
}

@article{Hosseini2016,
  title = {Partially Nondestructive Continuous Detection of Individual Traveling Optical Photons},
  author = {Mahdi Hosseini and Kristin M. Beck and Yiheng Duan and Wenlan Chen and Vladan Vuleti{\'c}},
  journal = {Physical Review Letters},
  volume = {116},
  number = {3},
  pages = {033602},
  year = {2016},
  publisher = {American Physical Society},
  doi = {10.1103/PhysRevLett.116.033602}
}

@article{Hallaji2017,
  title = {Weak-value amplification of the nonlinear effect of a single photon},
  author = {Matin Hallaji and Amir Feizpour and Greg Dmochowski and Josiah Sinclair and Aephraim M. Steinberg},
  journal = {Nature Physics},
  volume = {13},
  number = {6},
  pages = {540--544},
  year = {2017},
  publisher = {Nature Publishing Group},
  doi = {10.1038/nphys4040}
}

@article{Duan2020,
  title = {Heralded Interaction Control between Quantum Systems},
  author = {Yiheng Duan and Mahdi Hosseini and Kristin M. Beck and Vladan Vuleti{\'c}},
  journal = {Physical Review Letters},
  volume = {124},
  number = {22},
  pages = {223602},
  year = {2020},
  publisher = {American Physical Society},
  doi = {10.1103/PhysRevLett.124.223602}
}

@article{Milburn1989,
  title = {Quantum optical Fredkin gate},
  author = {G. J. Milburn},
  journal = {Physical Review Letters},
  volume = {62},
  number = {18},
  pages = {2124--2127},
  year = {1989},
  publisher = {American Physical Society},
  doi = {10.1103/PhysRevLett.62.2124}
}

@article{Nemoto2004,
  title = {Nearly Deterministic Linear Optical Controlled-NOT Gate},
  author = {Kae Nemoto and W. J. Munro},
  journal = {Physical Review Letters},
  volume = {93},
  number = {25},
  pages = {250502},
  year = {2004},
  publisher = {American Physical Society},
  doi = {10.1103/PhysRevLett.93.250502}
}

@article{Feizpour2011,
  title = {Amplifying Single-Photon Nonlinearity Using Weak Measurements},
  author = {Amir Feizpour and Xingxing Xing and Aephraim M. Steinberg},
  journal = {Physical Review Letters},
  volume = {107},
  number = {13},
  pages = {133603},
  year = {2011},
  publisher = {American Physical Society},
  doi = {10.1103/PhysRevLett.107.133603}
}

@article{Sinclair2022,
  title = {Measuring the Time Atoms Spend in the Excited State Due to a Photon They Do Not Absorb},
  author = {Josiah Sinclair and Daniela Angulo and Kyle Thompson and Kent Bonsma-Fisher and Aharon Brodutch and Aephraim M. Steinberg},
  journal = {PRX Quantum},
  volume = {3},
  number = {1},
  pages = {010314},
  year = {2022},
  publisher = {American Physical Society},
  doi = {10.1103/PRXQuantum.3.010314}
}

@article{Angulo2026,
  title = {Experimental Observation of Negative Weak Values for the Time Atoms Spend in the Excited State as a Photon Is Transmitted},
  author = {Angulo, Daniela and Thompson, Kyle and Nixon, Vida-Michelle and Jiao, Andy and Wiseman, Howard M. and Steinberg, Aephraim M.},
  journal = {Physical Review Letters},
  volume = {136},
  issue = {15},
  pages = {153601},
  numpages = {7},
  year = {2026},
  month = {Apr},
  publisher = {American Physical Society},
  doi = {10.1103/gjfq-k9dv},
  url = {https://link.aps.org/doi/10.1103/gjfq-k9dv}
}

@article{Sondermann2007,
  author  = {M. Sondermann and R. Maiwald and H. Konermann and
             N. Lindlein and U. Peschel and G. Leuchs},
  title   = {Design of a mode converter for efficient light-atom coupling in free space},
  journal = {Applied Physics B},
  volume  = {89},
  number  = {4},
  pages   = {489--492},
  year    = {2007},
  doi     = {10.1007/s00340-007-2859-4}
}

@article{Wang2011,
  title = {Efficient excitation of a two-level atom by a single photon in a propagating mode},
  author = {Yimin Wang and Jiří Minář and Lana Sheridan and Valerio Scarani},
  journal = {Physical Review A},
  volume = {83},
  number = {6},
  pages = {063842},
  year = {2011},
  publisher = {American Physical Society},
  doi = {10.1103/PhysRevA.83.063842}
}

@article{Golla2012,
  author  = {A. Golla and B. Chalopin and M. Bader and I. Harder and
             K. Mantel and R. Maiwald and N. Lindlein and
             M. Sondermann and G. Leuchs},
  title   = {Generation of a wave packet tailored to efficient free space excitation of a single atom},
  journal = {European Physical Journal D},
  volume  = {66},
  pages   = {190},
  year    = {2012},
  doi     = {10.1140/epjd/e2012-30293-y}
}

@article{Aljunid2013,
  title = {Excitation of a single atom with exponentially rising light pulses},
  author = {Syed Abdullah Aljunid and Gleb Maslennikov and Yimin Wang and Hoang Lan Dao and Valerio Scarani and Christian Kurtsiefer},
  journal = {Physical Review Letters},
  volume = {111},
  number = {10},
  pages = {103001},
  year = {2013},
  publisher = {American Physical Society},
  doi = {10.1103/PhysRevLett.111.103001}
}

@article{Leong2016,
  author  = {Victor Leong and Mathias Alexander Seidler and Matthias Steiner and Alessandro Cer{\`e} and Christian Kurtsiefer},
  title   = {Time-resolved scattering of a single photon by a single atom},
  journal = {Nature Communications},
  volume   = {7},
  pages    = {13716},
  year     = {2016},
  doi      = {10.1038/ncomms13716},
  publisher = {Nature Publishing Group}
}

@article{Rag2017,
  title = {Two-level-atom excitation probability for single- and N-photon wave packets},
  author = {Hemlin Swaran Rag and Julio Gea-Banacloche},
  journal = {Physical Review A},
  volume = {96},
  number = {3},
  pages = {033817},
  year = {2017},
  publisher = {American Physical Society},
  doi = {10.1103/PhysRevA.96.033817}
}

@article{Wiseman2023,
  author       = {Wiseman, Howard M. and Steinberg, Aephraim M. and Hallaji, Matin},
  title        = {Obtaining a single-photon weak value from experiments using a strong (many-photon) coherent state},
  journal      = {AVS Quantum Science},
  volume       = {5},
  number       = {2},
  pages        = {024401},
  year         = {2023},
  doi          = {10.1116/5.0137579},
  url          = {https://doi.org/10.1116/5.0137579},
}

@article{Crisp1970,
  author  = {Crisp, M. D.},
  title   = {Propagation of Small-Area Pulses of Coherent Light through a Resonant Medium},
  journal = {Physical Review A},
  volume  = {1},
  number  = {6},
  pages   = {1604--1611},
  year    = {1970},
  doi     = {10.1103/PhysRevA.1.1604},
}

@article{Burnham1969,
  author  = {Burnham, David C. and Chiao, Raymond Y.},
  title   = {Coherent Resonance Fluorescence Excited by Short Light Pulses},
  journal = {Physical Review},
  volume  = {188},
  number  = {2},
  pages   = {667--675},
  year    = {1969},
  doi     = {10.1103/PhysRev.188.667},
}

@article{Carmichael2000,
  title        = {Giant Violations of Classical Inequalities through Conditional Homodyne Detection of the Quadrature Amplitudes of Light},
  author       = {Carmichael, H. J. and Castro-Beltran, H. M. and Foster, G. T. and Orozco, L. A.},
  journal      = {Physical Review Letters},
  volume       = {85},
  number       = {9},
  pages        = {1855--1858},
  year         = {2000},
  month        = aug,
  doi          = {10.1103/PhysRevLett.85.1855}
}

@article{Foster2000,
  author = {Foster, G. T. and Orozco, L. A. and Castro-Beltran, H. M. and Carmichael, H. J.},
  title = {Quantum State Reduction and Conditional Time Evolution of Wave-Particle Correlations in Cavity QED},
  journal = {Physical Review Letters},
  volume = {85},
  number = {15},
  pages = {3149--3152},
  year = {2000},
  publisher = {American Physical Society},
  doi = {10.1103/PhysRevLett.85.3149}
}

@article{Foster2002,
  author = {Foster, G. T. and Smith, W. P. and Reiner, J. E. and Orozco, L. A.},
  title = {Time-dependent electric field fluctuations at the subphoton level},
  journal = {Physical Review A},
  volume = {66},
  number = {3},
  pages = {033807},
  year = {2002},
  publisher = {American Physical Society},
  doi = {10.1103/PhysRevA.66.033807}
}

@article{Masters2023,
  title        = {On the simultaneous scattering of two photons by a single two-level atom},
  author       = {Masters, Luke and Hu, Xin-Xin and Cordier, Martin and Maron, Gabriele and Pache, Lucas and Rauschenbeutel, Arno and Schemmer, Max and Volz, J{\"u}rgen},
  journal      = {Nature Photonics},
  volume       = {17},
  number       = {11},
  pages        = {972--976},
  year         = {2023},
  month        = jul,
  doi          = {10.1038/s41566-023-01260-7}
}

@article{Wang2025,
  title        = {Purcell-Enhanced Generation of Photonic Bell States via the Inelastic Scattering off Single Atoms},
  author       = {Wang, Jian and Zhou, Xiao-Long and Shen, Ze-Min and Huang, Dong-Yu and He, Si-Jian and Huang, Qi-Yang and Liu, Yi-Jia and Li, Chuan-Feng and Guo, Guang-Can},
  journal      = {Physical Review Letters},
  volume       = {134},
  number       = {5},
  pages        = {053401},
  year         = {2025},
  month        = feb,
  doi          = {10.1103/PhysRevLett.134.053401}
}

\section{End Matter}

\subsection{An Analytic Derivation of the Atomic Excitation Weak Value}
The goal of this section is to give an analytic derivation of the time-dependent weak value of atomic excitation, and show the $1-\exp(OD/2)$ peak behavior. We will do so in the ideal situation where the photon is post-selected as a $\delta$-function in time.

The frequency responses of the photon amplitude given in Ref. \cite{Thompson2025} (written in terms of the optical depth, OD, of the atoms) are

\begin{equation}
\alpha'_{\rightarrow, \leftarrow}(\omega)=\alpha_{\rightarrow, \leftarrow}(\omega)e^{-\frac{OD}{2} \frac{\Gamma/2}{-i\omega\pm\Gamma/2}}
\label{eq:eqs_of_motion}
\end{equation}

We assume the medium is infinitely thin. $\rightarrow$ is for the forward-evolving (initial) state. $\leftarrow$ is for the backward-evolving (final) state. 

To find the weak value for atomic excitation, we use the relation

\begin{equation}
P_{\gamma}(t) + P_e(t) = 1
\label{eq:conservation}
\end{equation}

$P_\gamma$ and $P_e$ are respectively values of the projection operators for being a photon and for being an atomic excitation.

Their weak values are given by

$$
P_{\gamma,e}(t) = \frac{\bra{f(t)}\hat\Pi_{\gamma,e}\ket{i(t)}}{\langle f(t)|i(t)\rangle}
$$

\begin{figure}[h]
    \centering
    \includegraphics[width = 0.8\linewidth]{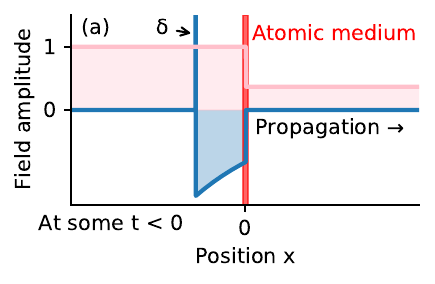}
    \includegraphics[width = 0.8\linewidth]{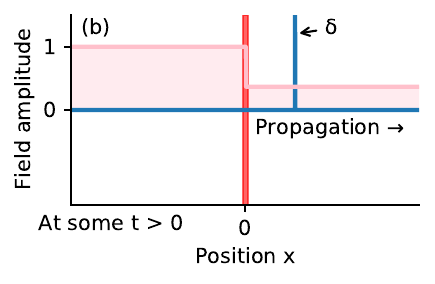}
    \caption{Illustrations of the forward evolving state (pink) and the backward evolving state (blue) for times (a) before the retrodicted wavepacket entered the atomic medium and (b) after the wavepacket exited and is to be detected. The wavepacket of the backward-evolving state exits at t = 0. The atomic medium (red) is infinitely thin and is positioned at x = 0.}
    \label{fig:derivation}
\end{figure}

Describing the forward evolving state in the time domain, $\alpha_{\rightarrow}(t) = 1$ before the atom cloud and $\alpha_{\rightarrow} = e^{-\frac{OD}{2}}$ after the atom cloud.

Describing the backward evolving state, it is a $\delta$-function after the atom cloud.

The solution to eq.\ref{eq:eqs_of_motion} of the backward evolving state (for $t < 0$) before the atom cloud is

$$
G^{-1}_B(t) = \delta(t) - \frac{OD\Gamma}{2}\frac{J_1(\sqrt{-OD\Gamma t})}{\sqrt{-OD\Gamma t}}e^{\frac{\Gamma}{2} t}
$$

One can show that $G^{-1}_B(t) = G(-t)$, and $G(t)$ is given in the literature \cite{Burnham1969, Crisp1970}. $J_1$ is the Bessel function of the first kind.

$G(t)$ is the Green's function of the frequency response for the forward evolving state in eq. \ref{eq:eqs_of_motion}. $G^{-1}_B(t)$ is what the input is if the output is $\delta(t)$ for the backward evolving state.

The inner product between the forward and backward evolving states does not change in time and is $e^{-OD/2}$.

Since the forward evolving state is constant before the atoms, one integrates $G_B^{-1}(t')$ from $t'=t$ to $t'=0$ to evaluate $P_{\gamma}(t)$ (for $t<0$).

The $\delta$-function in $G_B^{-1}(t')$ will be included in the integration for all $t<0$. It is convenient to separate it out in the integration.

In equations,

\begin{align*}
P_{\gamma}(t) &= \frac{\int_{t}^{0}G^{-1}_B(t')dt'}{e^{-\frac{OD}{2}}} \\
&=e^{\frac{OD}{2}}\left( 1 - \frac{OD\Gamma}{2}\int_{t}^{0}{\frac{J_1(\sqrt{-OD\Gamma t'})}{\sqrt{-OD\Gamma t'}}e^{\frac{\Gamma}{2} t'}dt'}\right)
\end{align*}

Applying this to eq.\ref{eq:conservation} gives the desired result,

\begin{equation}
P_e(t) = 1 - e^{\frac{OD}{2}} + e^{\frac{OD}{2}}\frac{OD\Gamma}{2}\int_{t}^{0}{\frac{J_1(\sqrt{-OD\Gamma t'})}{\sqrt{-OD\Gamma t'}}e^{\frac{\Gamma}{2} t'}dt'}
\end{equation}

(The above analytic expression when evaluated using numerical integration agrees with the simulation solving the equations of motion.)

Final remarks: At $t = 0$, which is when the peak occurs,

$$
P_e = 1-e^{\frac{OD}{2}}
$$

Thus, at high ODs, the peak scales exponentially with OD.

The exponential scaling, a result of the inner product of the forward and backward evolving states being $e^{-OD/2}$, is an example of weak value amplification.

\subsection{Obtaining the Single Photon Weak Value in this Experiment}
For a coherent-state input and a given detection mode, the difference between the weak values conditioned on photon detection and no detection equals the weak value of a single photon prepared in the input mode and post-selected in the detection mode. 

In the context of this experiment, the no detection weak value (with no photon detected anywhere in the measurement window) is flat in time. Subtracting off a linear fit in the data analysis effectively removes this no detection weak value, since the sparse photons detected do not affect the flat background.

\begin{figure}[h]
    \centering
    \includegraphics[width = \linewidth]{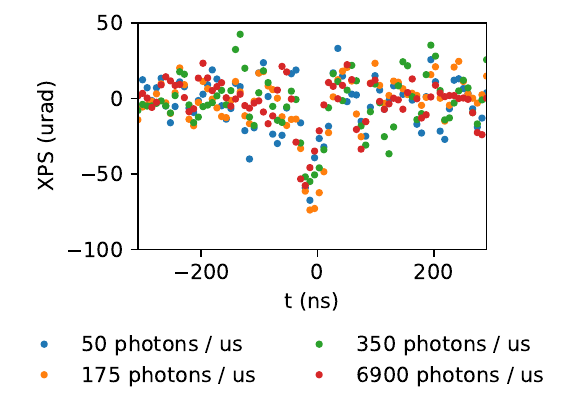}
    \caption{Data taken at OD = 2.5$\pm$0.2 for different signal beam powers. The input powers of the signal beam are expressed in terms of photon rates, given in the legend. }
    \label{fig:vs_intensity}
\end{figure}

A feature of the technique is that the measurement does not depend on the intensity of the coherent state input. We took data using multiple signal beam powers at the same OD for corroboration, as shown in Fig. \ref{fig:vs_intensity}.

Given a beam waist radius of 25 um, the maximum intensity (given in photons per absorption cross section per lifetime) is the power multiplied by $7.7\times10^{-6}\mu m^2\cdot \mu s$. (The absorption cross section used is for the $m_F = 3$ to $m_F = 4$ transition, which is 0.29 $\mu m^2$.) Thus, the signal beam with the highest power in the data produces a maximum intensity of 0.05 photons per absorption cross section per lifetime, far below saturation. The saturation intensity is $1/2$ photon per absorption cross section per lifetime.  

\end{document}